# The Structure of Weighting Coefficient Matrices of Harmonic Differential Quadrature and Its Applications

WEN CHEN[a], XINWEI WANG[b] AND TINGXIU ZHONG[a]

[a]Department of Mechanical Engineering, Shanghai Jiao Tong University, Shanghai 200030, P.R.China.

[b]Department of Aircraft Engineering, Nanjing University of Aeronautics and Astronautics, Nanjing 210016, P.R.China.

**KEY WORDS**: harmonic differential quadrature method, matrix, differential quadrature, numerical method, centrosymmetric matrix.

## SUMMARY

The structure of weighting coefficient matrices of Harmonic Differential Quadrature (HDQ) is found to be either centrosymmetric or skew centrosymmetric depending on the order of the corresponding derivatives. The properties of both matrices are briefly discussed in this paper. It is noted that the computational effort of the harmonic quadrature for some problems can be further reduced up to 75 per cent by using the properties of the above-mentioned matrices.

# 1. INTRODUCTION

Harmonic differential quadrature [1, 2] is a new development of the differential quadrature (DQ) [3-6], which has been used successfully to solve a variety of problems. Harmonic differential quadrature chooses harmonic functions as its test functions instead of polynomials as in the differential quadrature and has been found especially efficient for problems with periodic behaviors [1, 2]. Chen and Yu [7] pointed out earlier that the weighting coefficient matrices of DQ method possess centrosymmetric and skew centrosymmetric structure if the grid spacing is symmetric. And a new type matrix, called the skew centrosymmetric matrix, was introduced in that paper.

In the present study, it is found that the weighting coefficient matrices of HDQ has similar structure to the DQ method if the grid spacing is symmetric. It is known that the properties of centrosymmetric matrices can be used to factorize its determinant and characteristic equations into two smaller size sub-matrices [8]. Therefore, the computational effort of the HDQ method can be further reduced up to 75 per cent by using the properties of the centrosymmetric and skew centrosymmetric matrices. The free vibration of rectangular plate is investigated to show the efficiency of the HDQ method by using the centrosymmetric behavior. For completeness, we also discuss briefly the behaviors of both the centrosymmetric and skew centrosymmetric matrices and found that they possess similar properties. Finally, some conclusions are drawn based on the results reported herein.

## 2. CENTROSYMMETRIC AND SKEW CENTROSYMMETRIC MATRICES

The following states some known properties of centrosymmetric matrix only for odd order without proofs [8], since the HDQ method usually uses the odd number of grid points. The structural properties of skew centrosymmetric matrix is analyzed. Proofs for skew centrosymmetric matrices are similar as those for centrosymmetric matrix, so they are omitted herein for brevity.

Definition 1. The contra-identity matrix J has ones along the secondary diagonal and zeros elsewhere. Note that $J^T = J$ and $J^2 = I$, where I is the unit matrix. The effect of premultiplying in matrix by J reverses the order of its rows, and postmultiplying reverses the order of its columns.

Definition 2. An N dimensional vector $\gamma$ is defined to be symmetric if

$$J\gamma = \gamma \tag{1}$$

or skew symmetric if

$$J\gamma = -\gamma . \tag{2}$$

Definition 3. Let $C_{N \times N}$ denotes the set of N×N real Centrosymmetric matrix, then

$$X \in C_{N \times N}, \text{ if and only if } JXJ = X. \tag{3}$$

Definition 4. Let $NC_{N \times N}$ denotes the set of N×N real skew centrosymmetric matrices. then,

$$X \in C_{N \times N}, \text{ if and only if } JXJ = -X. \tag{4}$$

Lemma 1. If $X \in C_{N \times N}$, $Y \in C_{N \times N}$, $Z \in NC_{N \times N}$, then $XY \in C_{N \times N}$, $X+Y \in C_{N \times N}$, $XZ \in NC_{N \times N}$, $X^{-1} \in C_{N \times N}$, and $Z^{-1} \in NC_{N \times N}$.

Lemma 2. If N is odd (N=2M+1), any matrix $Q \in C_{N \times N}$ can be written as

$$Q = \begin{bmatrix} A & Js & JCJ \\ t & q & tJ \\ C & s & JAJ \end{bmatrix} \qquad (5)$$

where A and C are arbitrary M×M matrices, s and t are vectors of M×1 dimension, and q is a scalar. It is easily proved that

$$Q = \begin{bmatrix} A & Js & JCJ \\ t & q & tJ \\ C & s & JAJ \end{bmatrix} \text{ and } \begin{bmatrix} A - JC & 0 & 0 \\ 0 & q & 2t \\ 0 & 2s & A + JC \end{bmatrix} \qquad (6)$$

are orthogonally similar. Thus, the evaluation of eigenvectors and eigenvalues of Q is equivalent to those of both M×M and (M+1)×(M+1) matrices.

Lemma 3. The eigenvector of centrosymmetric matrix Q in Lemma 3 is the sum of a symmetric vector and a skew symmetric vectors. If Q has distinct eigenvalues and N is odd number (2M+1), the M skew symmetric orthonormal eigenvectors $U_i$ and corresponding eigenvalues can be determined by

$$(A - JC)x_i = \lambda_i x_i \qquad (7)$$

where $U_i = [x_i^T, 0, -(Jx_i^T)^T]^T$. The M+1 symmetric eigenvectors $V_i$ and corresponding eigenvalues $\rho_i$ are obtained by solving the following equation

$$\begin{bmatrix} q & 2t \\ 2s & A+JC \end{bmatrix} \begin{Bmatrix} \alpha_i \\ y_i \end{Bmatrix} = \rho_i \begin{Bmatrix} \alpha_i \\ y_i \end{Bmatrix} \tag{8}$$

where $V_i = [y_i^T, 2\alpha_i, (Jy_i^T)^T]^T$.

Lemma 4. If $Q \in C_{N \times N}$ and $N=2M+1$, Q can be restated as

$$Q = \begin{bmatrix} A & -Js & -JCJ \\ t & 0 & -tJ \\ C & s & JAJ \end{bmatrix} \tag{9}$$

where A and C are arbitrary M×M matrices, s and t are vectors of M×1 vector. We can prove that

$$Q = \begin{bmatrix} A & -Js & -JCJ \\ t & 0 & -tJ \\ C & s & JAJ \end{bmatrix} \text{ and } \begin{bmatrix} 0 & -2s & A-JC \\ 2t & 0 & 0 \\ A+JC & 0 & 0 \end{bmatrix} \tag{10}$$

are orthogonally similar. Thus, the evaluation of eigenvectors and eigenvalues of Q is equivalent to that of a (M+1)×(M+1) matrix.

As can be observed from the foregoing Lemmas. the calculation of the eigenvalues and eigenvectors of a centrosymmetric matrix be reduced to those of two smaller matrices, which expedites the computational effort by nearly 75%. For skew centrosymmetric matrices, we find that there are similar computational reduction effects for the eigenvalue and eigenvectors. Note that premultiplying or postmultiplying any matrix by J only moves the elements of matrix and consumes very little computation time.

## 3. THE CENTROSYMMETRIC AND SKEW CENTROSYMMETRIC STRUCTURE OF WEIGHTING COEFFICIENT MATRICES IN THE HDQ METHOD

The essence of the harmonic differential quadrature method is that the derivative of a function, with respect to a variable at a given discrete point, is approximated as a weighted linear sum of the function values at all discrete points. Consider a function f(x) of one variable in the domain (0,1) with N discrete grid points, a harmonic differential quadrature approximation at the ith discrete point is given by

$$\frac{\partial^m f(x)}{\partial x^m}\bigg|_{x=x_i} = \sum_{j=1}^{N} w_{ij} f(x_j), \qquad i = 1, 2, \ldots, N. \qquad (11)$$

where $x_j$ are the discrete points in the variable domain. $f(x_j)$ and $w_{ij}^{(m)}$ are the function values at these points and the weighting coefficients, respectively. The $w_{ij}^{(m)}$ are obtained by requiring that Eq. (11) be exact at all grid points. Unlike differential quadrature method, harmonic differential quadrature uses harmonic functions as its test functions, i.e.,

$$f(x) = \{1, \sin\pi x, \cos\pi x, \sin 2\pi x, \cos 2\pi x, \ldots, \sin(N-1)\pi x/2, \cos(N-1)\pi x/2\}$$

(12)

where N is an odd number. Substituting Eq. (12) into Eq. (11), we obtain N sets of N order algebraic equations. In matrix form, we have

$$H w^{(m)} = H^{(m)} \qquad (13)$$

where matrix $w^{(m)}$ denotes N×N weighting coefficient matrix. m is the order of the derivative, matrix H is determined by the harmonic test functions (12) and the grid spacing. Matrix $H^{(m)}$ depends on the test functions, the grid spacing and the derivatives order m. Premultiplying Eq. (13) by $H^T$ yields

$$H^T H w^{(m)} = H^T H^{(m)} \qquad (14)$$

The grid spacings usually used are symmetric such as equally spaced grid points and unequally spaced ones adopting the roots of the shifted Legendre or Chebyshev polynomials. If we choose symmetric grid spacings, i.e.,

$$x_{N+1-k} = 1 - X_k, \quad k = 1, 2, \ldots, N. \qquad (15)$$

it can easily proven that the product $H^T H$ is centrosymmetric, and $H^T H^{(m)}$ is centrosymmetric if m is even number or skew centrosymmetric if m is odd number. From Eq. (14), we have

$$w^{(m)} = (H^T H)^{-1} H^T H^{(m)} \qquad (16)$$

According to Lemma 1, $(H^T H)^{-1} \in C_{N \times N}$. Thus, the weighting coefficient matrices $w^{(m)}$ are centrosymmetric for even order derivatives and skew centrosymmetric for odd order derivatives.

## 4. EXAMPLE

The approach for applying boundary conditions presented by Wang and Bert [6] is used in this paper. According to the method [6], the weighting coefficient matrices will be modified in terms of boundary conditions. It is straightforward that such coefficient matrices are also centrosymmetric if the boundary conditions are symmetric. Consider the transverse vibration of a thin rectangular SS-SS-SS-SS plate and SS-C-SS-C plate, where C and SS denote clamped and simply supported boundary conditions, respectively. Let $\bar{\omega}^2 = \rho h a^4 \omega^2 / D$, $\alpha = a/b$ and a is the width and b length of plate. In terms of the harmonic differential quadrature, the governing equation can be written as (the details see references [2, 6])

$$\sum_{k=2}^{N_x-1} \bar{D}_{ik} W_{kj} + (2\alpha^2) \sum_{m=2}^{N_y-1} \bar{B}_{jm} \sum_{k=2}^{N_x-1} \bar{B}_{ik} W_{km} + (\alpha^4) \sum_{k=2}^{N_y-1} \bar{D}_{jk} W_{ik} = \bar{\omega}^2 W_{ij} \quad (17)$$

i,j =2, 3, ..., N-1.

where $\bar{B}$ and $\bar{D}$ are the modified HDQ weighting coefficients for the second and fourth order derivatives, respectively. $N_x$ and $N_y$ are the number of grid points in the X- and Y- directions (Cartesian coordinates). Equally spaced grid point is used for the present cases. But for plate with C-C boundary conditions, tow grid points, separated by a small distance, are placed at each boundary and uniform grid spacing is used for the inner grid points.

HDQ with a rather small number of grid points can produces very good result. For square SS-SS-SS-SS plate, five grid points are used and the HDQ results for the fundamental frequency of $\bar{\omega} = \pi$ and for the 2nd frequency of $\bar{\omega} = 2\pi$ are coincident with exact solutions. For SS-C-SS-C plate, we use eleven grid points and list some HDQ results in table 1 which are coincident with HDQ solutions [2] and very close to those of Leissa [9], while the computational effort in this paper is reduced to only about 1.6% as much as that in reference [2]. This computational savings is achieved by using centrosymmetric properties of the HDQ weighting coefficient matrices (Lemmas 2 and 3 in section 2). We computes only a $(N^2-2)/4 \times (N^2-2)/4$ matrix for the present case. In contrast, references [2] need to solve a $(N^2-2) \times (N^2-2)$ matrix. But it is regret that the reduced method is only available for problems with systemic boundary conditions. For all other examples with sysmetic boundary given in reference [2], we obtain the same results by this reduced HDQ method.

It is known that when more grid points is used, the accuracies of HDQ results can be improved. Unlike the DQ method, the HDQ method have not limitation for the number of grid points. But for more grid points computing effort increases rapidly especially for multi-dimensional problems. Thus, the present reduced method will be significant in practice. In subsequent papers, we will give more complex examples by using the present reduced HDQ method to show its affective computational savings.

## 5. CONCLUSION

The present work furthers the knowledge and understanding of harmonic differential quadrature. The primary contributions of this research are of finding the centrosymmetric and skew centrosymmetric structure of weighting coefficient matrices of harmonic differential quadrature and of discussing some properties of a new type matrix, skew centrosymmetric matrix. The centrosymmetric and skew centrosymmetric matrices have factorization properties which reduce significantly the computational complexity of eigenvalue and eigenvector calculations. Therefore, the present work makes HDQ more efficient for practical engineering application.

**Table 1.** Fundamental frequency ($\varpi^2 = \rho h a^4 \omega^2 / D$) for SS-C-SS-C plates

| a/b    | 0.4     | 2/3     | 1.0     | 1.5     | 2.5     |
|--------|---------|---------|---------|---------|---------|
| HDQ    | 12.1408 | 17.3821 | 28.9656 | 56.3752 | 145.550 |
| Leissa | 12.1347 | 17.3703 | 28.9509 | 56.3481 | 145.484 |